# A Practical Approach for Detecting Logical Error in Object Oriented Environment

Ghassan Samara
Internet Technology Department
Zarqa University, Jordan.

Abstract— A programming language is a formally constructed language designed to communicate instructions to a machine, particularly a computer. Programming languages can be used to create programs to control the behavior of a machine or to express algorithms. Most programs that are written by programmers are often compiled correctly with no syntax or semantic errors. However, some other errors appear after the execution of the program (logical error).

Logical Errors (LE) are errors that remain after all syntax errors have been removed. Usually, the compiler does not detect LE, so the produced results are different from what the programmer is expecting. For this reason, discovering and fixing the logical error is very hard and proposes a good topic for research and practice.

Some LE are resulted from the misuse of classes' objects, and in Software Development Life Cycle (SDLC), it is considered that the software with LE is low-quality software with high maintenance cost.

In this paper, an object-oriented environment that allows the programmer to detect and discover LE to avoid it. This environment will be called Object Behavior Environment (OBEnvironment) will enforce the correct use of objects according to their predefined behaviors by using tools like Xceed Component (that appeal .Net windows form developers for building better applications), Alsing Component (that provides an area to programmer that allows writing correct syntax code by C# language) and, Mind Fusion Component (that provides an area to programmer that allows drawing State Diagrams to show object state).

## I. INTRODUCTION

LE is one of the major problems in software testing that appears in the last stages of the SDLC. So, a tool is needed to improve the compilers of the Object Oriented (OO) languages, and this could be done by detecting logical errors before the executing the software product.

In this research, a C# compiler will be improved to detect these errors before compilation by controlling and tracing the object behavior of the objects in the application.

There are some weaknesses in the OO programming languages (for example C#) that it cannot detect LE before the execution time, for example, the misuse of objects in the application.

So finding these errors and correcting them will consume long time and high maintenance cost, so there should be a mechanism or tool to avoid this kind of errors.

### *1.1 The Object Behavior Environment*

The OBEnvironment uses source code (OO Application) and objects behaviors definitions for detecting logical errors; object behavior error is the errors related to the misuse of objects in the application.

It is assumed that the code used by the programmer is free of any syntax or semantic errors from existing files and draw a correct behavior model (State Diagram) of each used class according to their application.

Depending on the case type different State diagram for same source code could be drawn.

### *1.2 Object-Oriented Programming*

Object-oriented programming group's behavior (operations on data) and state (data) into modular units called objects and enables the programmer to combine objects into structured networks to form a complete program. Moreover, programmers can create relationships between one object and another [1].

One of the principal advantages of object-oriented programming techniques over procedural programming techniques is that it enables programmers to create modules that with no need to be changed when a new type of object is added. A programmer can simply create a new object that inherits many of its features from existing objects [2], this makes object-oriented programs easier to modify and increases reusability, modularization and software simplicity [3].

*1.3 Programming Errors*

Programmers do different types of errors when writing a program; even the most experienced programmers make mistakes. Knowing how to debug an application and to find errors is an important part of programming, there are three types of errors will be shown in the following:

*1.3.1 Compilation Errors*

Compilation errors, also known as compiler errors. When a syntax error occurred in the code, the compiler will detect the error and the program and the compilation will stop. At this point, the program will not be executed until the error is fixed [4].

For example, forgetting a semi-colon at the end of a statement, missing necessary punctuation, or try to use an (End If) statement without first using an (If statement). These errors can be detected by the compiler [4].

*1.3.2 Run Time Errors*

Run-time errors are errors that occur while a program is running. This typically occurs when the program attempts an operation that is impossible to execute, but this kind of errors cannot be detected until the running time.

An example of this is division by zero, or infinite loop [4].

*1.3.3 Logic Errors*

Logic errors are those errors that remain after all syntax errors have been removed. Usually, the compiler doesn't detect logic errors, so the programmer discovers the result of the program doesn't match the expected result. For this reason, finding this kind of errors is very hard.

The proposed project will provide an easy and strong environment that will help in detecting logical errors. These logical errors relate to misuse of objects. The proposed solution should be used after the implementation phase of developing any software product.

OBEnvironment will enforce the correct use of objects according to their defined behaviors.

To make the programming process easier, find errors that can't be detected.

To the Detect Object Behavior an easy to use and learn environment was created.

The Environment contains two tabs, in the first tab "liasing package" component is used which allows writing syntax code using c# language. In the second tab "MindFusion" component is used. It provides a panel for the programmer to draw State Diagram to show object state(s).

The interface used the .Net environment to produce familiar menus and commands like file, edit, trace and so on.

The interface enables the programmer to choose tracing object by object which can detect object behavior error only on the selected object, or tracing all objects which can detect all object behavior errors, and this gives more flexibility to the programmers.

## II. THE PROPOSED SYSTEM

To detect the object behavior an easy to use learning environment was created, the proposed environment contains two tabs, in the first tab a component called Aliasing package was used. Which provides an area to the programmer to write c# language code, and it cannot detect object behavior error.

In the second tab another component is used which is called MindFusion. This component provides an area to draw State Diagram to show object state(s).

The interface used in the proposed environment contains a .Net interface, such as file, edit, trace and so on.

*2.1 Experimental Setup*

*2.1.1 Programming Languages Used:*

*2.1.1.1 Net Environment*

- The .Net Framework is a completely new application development platform, several .Net products and various applications from Microsoft based on the .Net Framework, including a new version of exchange and SQL server, which are Extensible Markup Language (XML) enabled and integrated into the .Net platform.

- Several .Net services provided by Microsoft for use in developing applications running under the.Net Framework [8].

*2.1.1.2 The C# language*

The C# language is disarming simplicity with only about 80 keywords and a dozen built-in data types, but C# is highly expressive when it comes to implementing modern programming concepts. C# includes all the support for structured, component-based, object-oriented programming that one expects of a modern language built on the shoulders of C++ and Java.

Defining and working with classes is the heart of any object-oriented language. Classes define new types, allowing the programmer to extend the language to better model of the problem being solved. C# contains keywords for declaring new classes and their methods and properties, and for implementing encapsulation, inheritance, and polymorphism, the three pillars of object-oriented programming [9 and 10].

*Why C# .Net?*

C#. NET is robust, easy to use solution for developing complex project. Based on C, C++ and Java, programs that can be accessed by any one, allows communicating with different computer language, integrated design environment, makes programming and debugging fast and easy, rapid application development, and produce small size projects [9] [10].

*2.1.1.3 The Rational Rose Tools*

The Rational Rose was employed to draw an easy to read and understand UML diagrams. The rational rose is used for Use Case diagrams, Class diagrams, Sequence diagrams, and Activity diagrams.

*2.2 OBEnvironment Requirement*

Problem Specification

Stages to implement the OBEnvironment:

1-First Stage: Open an existing source code.

2-Second Stage: Drawing state diagram for the class or open an existing one and modify it.

3-Third Stage: Reorganize source code in a pattern for tracing purposes.

4-Fourth Stage: Transform of the state diagram into specification.

5-Fifth Stage: Compare between the application and the specification to check whether if the object were correctly used in the application to avoid some object behavior errors.

Detailed Specification

1. Open:

The programmer may open an existing source code and existing state diagram and edit it.

2. 2. Save:

The programmer can save the project (source code + state diagram).

3. Exit:

The programmer can close the project at any time with or without saving.

1. First Stage

1. Entering the project: The programmer can open an existing source code from a file.

2. Editing source code: The OBEnvironment allows the programmer to edit the source code. In this stage, there is a text area in the OBEnvironment.

2. Second Stage

Drawing state diagram: The OBEnvironment provides a graphical editor for the programmer to draw a state diagram (states and transactions between them) for the class or open existing one and modify it.

For example, see figure 1.

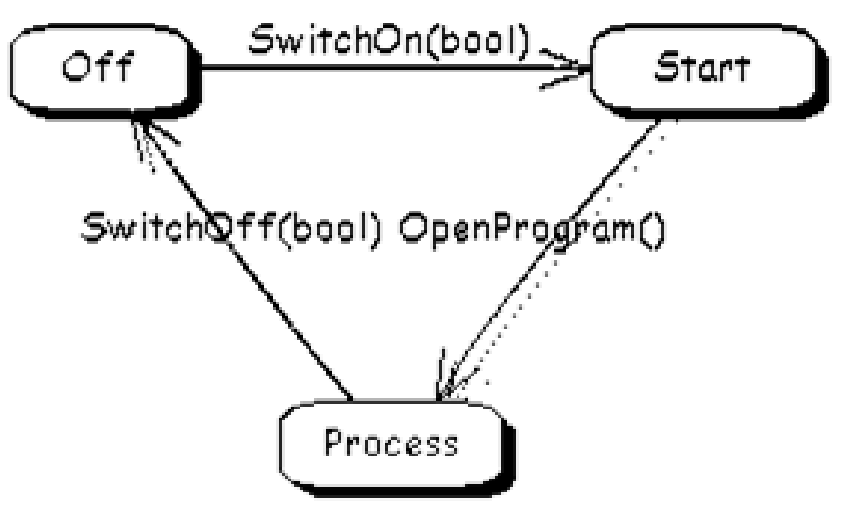

Figure 1: State diagram example.

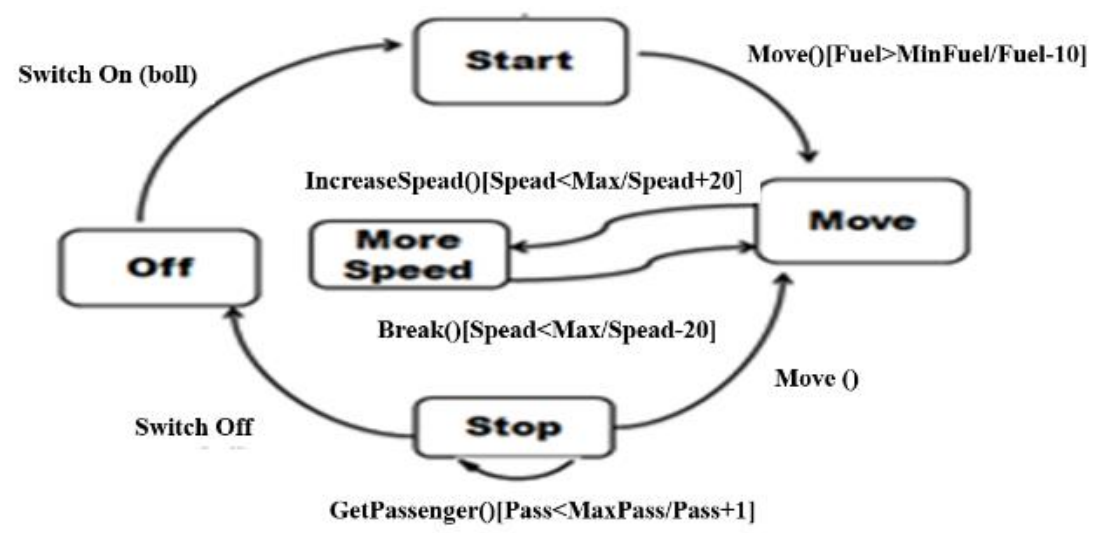

Figure 2: State diagram example.

3. Third Stage

Reorganize code: The OBEnvironment reorganizes source code into a pattern to extract the objects information for the comparison purposes.

4. Fourth Stage

Transformation: The OBEnvironment transforms the state diagram into a specification to extract the object behavior for comparing purposes. For example, see figure 3.

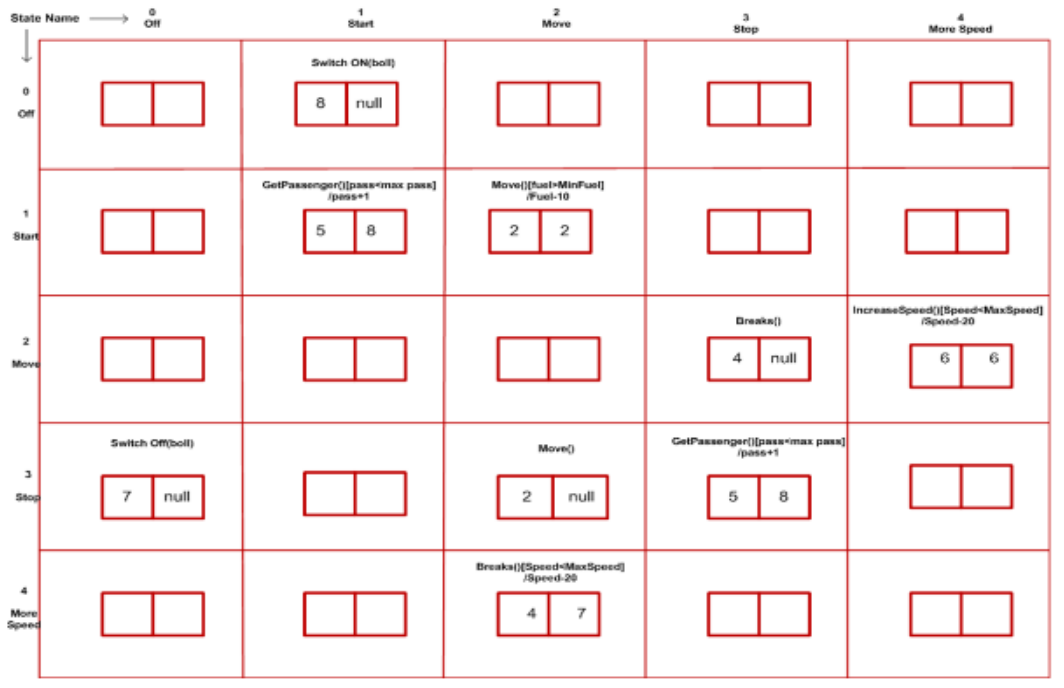

Figure 3: Transforming state diagram into specification Example.

5. Fifth Stage

Tracing: The OBEnvironment compare the specification with the use of objects in the application to detect some logical errors "object oriented violation".

*2.3 Object Behavior Environment Analysis*

*2.3.1 Interface Objects*

The OBEnvironment involves the following interface objects:

- Interface object "Open Project": This interface allows the programmer to interact with the system to open source code form existing file and edit it.

- Interface object "Drawing": This interface allows the programmer to draw a state diagram for a class to help the tracing process.

- Interface Object "Tracing ": This interface allows the programmer to transform state diagram into specification and use it to trace the object behavior.

- Interface Object "Showing Report ": This interface is presented to show a report to the programmer to help in locating the error if existed.

These interfaces are shown in the figure 4 below:

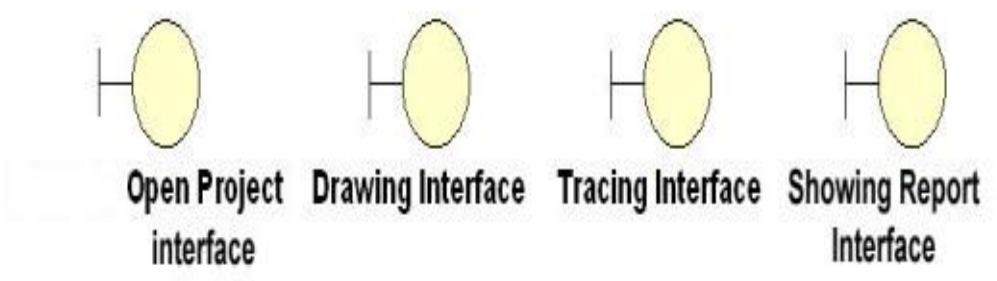

Figure 4: Interface Objects.

*2.3.2 Entity Objects*

OBEnvironment involves the following entity objects:

- Source code files: This entity object stores the information about the project's source code files.

-Draw: This entity object allows the programmer to store and retrieve the information about the state diagrams.

- Transformation: This entity allows the OBEnvironment to store and retrieve the needed specification for comparison purposes.

- Trace: This entity saves the result of tracing the object behavior to show it as a report to the programmer.

Figure 5 shows the four entities:

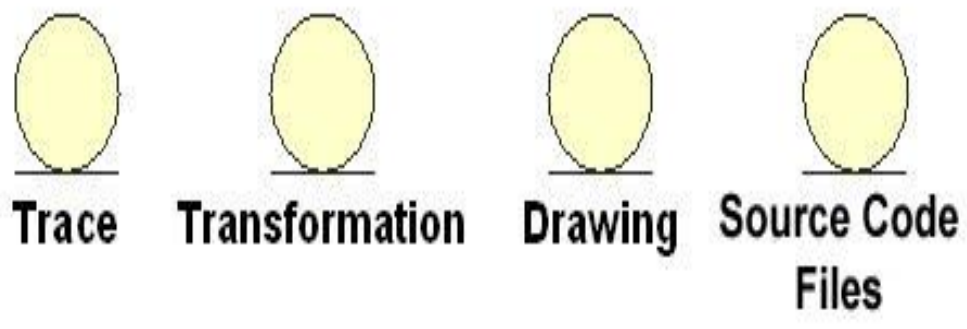

Figure 5: Entity Objects.

*2.3.3 Control Objects*

The OBEnvironment involves the following control objects:

- Source Code handler: This control object handles the programmer request to open a source code from an existing file.

- Drawer: This control object handles the programmer request to draw a state diagram.

- Transformer: This control object handles the programmers request in transforming the state diagram into specification.

- Tracer: This control object handles the programmers request in tracing the object behavior.

Figure 6 shows the four control objects:

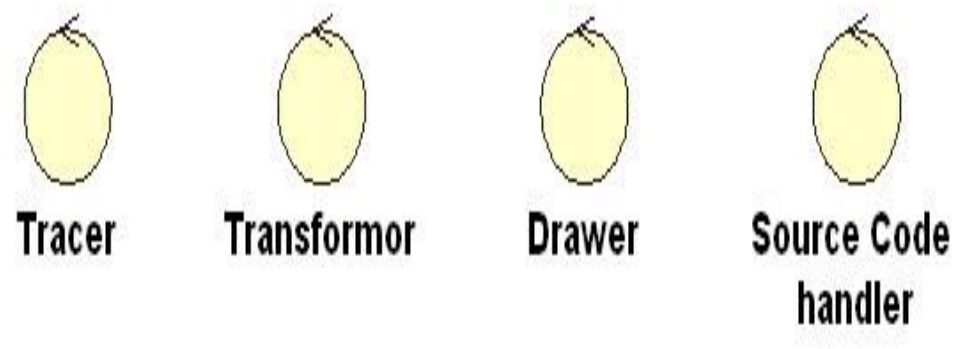

Figure 6: Control Objects.

The analysis model for the OBEnvironment is depicted in the figure 7:

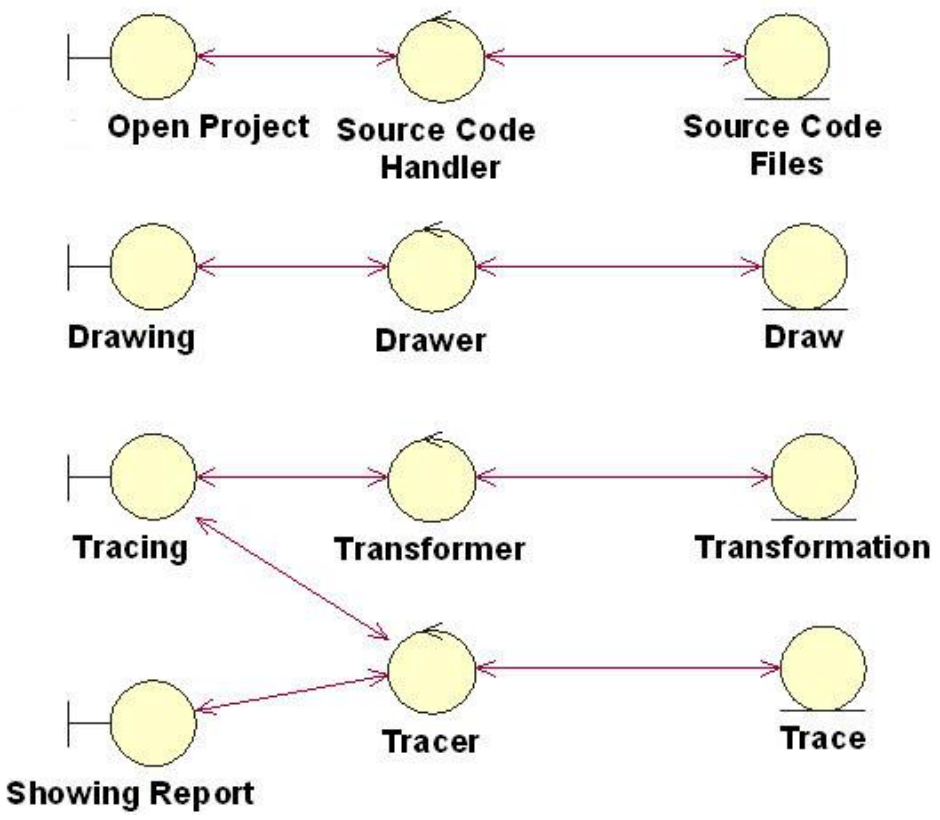

Figure 7: Analysis model for OBEnvironment.

*2.4 Subsystems*

Identify each subsystem individually:

*2.4.1 Source code files*

This subsystem starts when a programmer open a source code. The subsystem is shown in the figure 8:

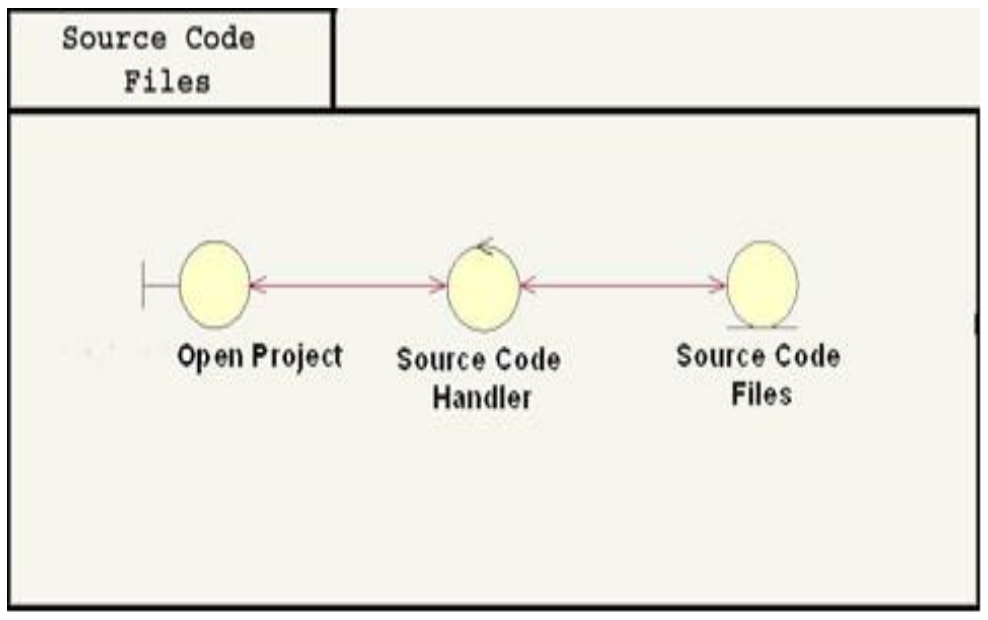

Figure 8: Source code files subsystem.

*2.4.2 Draw*

This subsystem starts when a programmer draws a state diagram. This subsystem is shown in the figure 9:

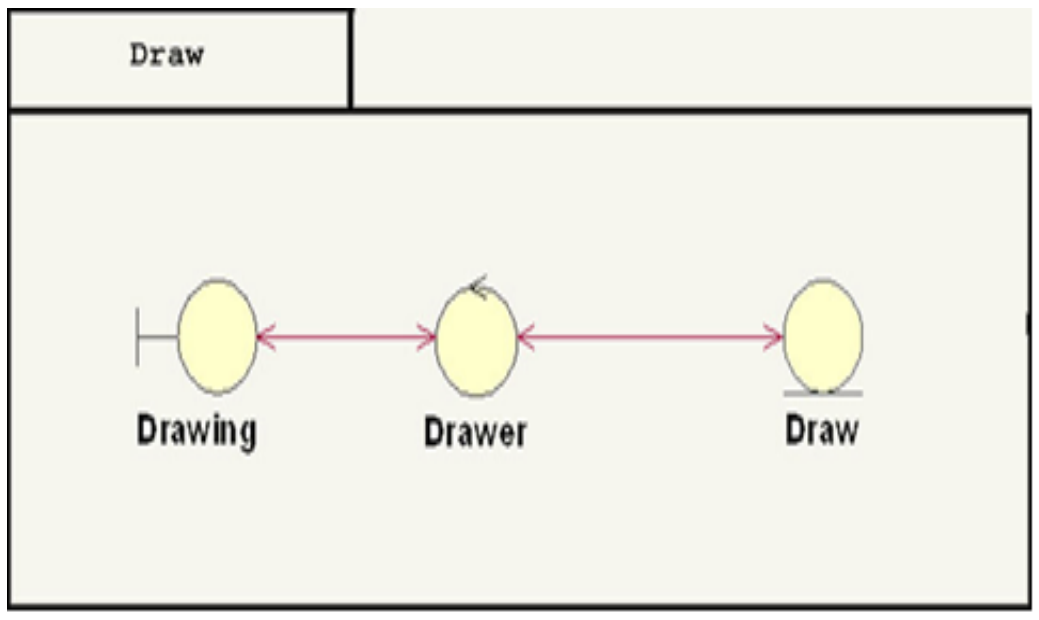

Figure 9: Draw subsystem.

*2.4.3 Transformation*

This subsystem starts when a programmer wants to transform the state diagram into specification. The subsystem is shown in the figure 10:

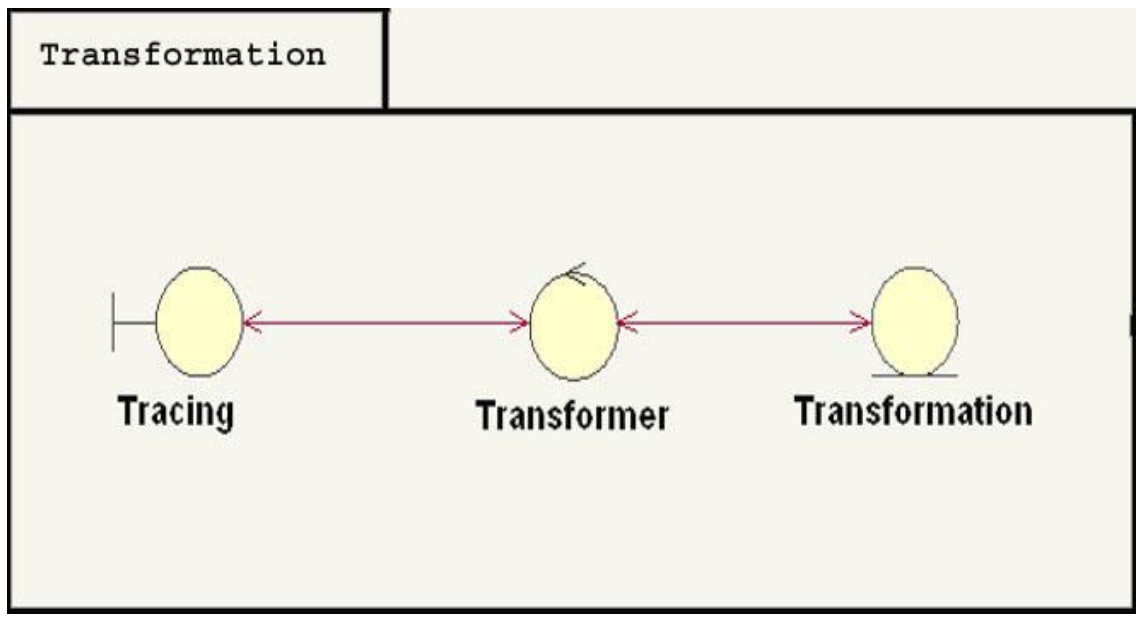

Figure 10: Transformation subsystem.

*2.4.4 Tracing*

This subsystem starts when a programmer traces the object behavior.

The subsystem is shown in the figure 11:

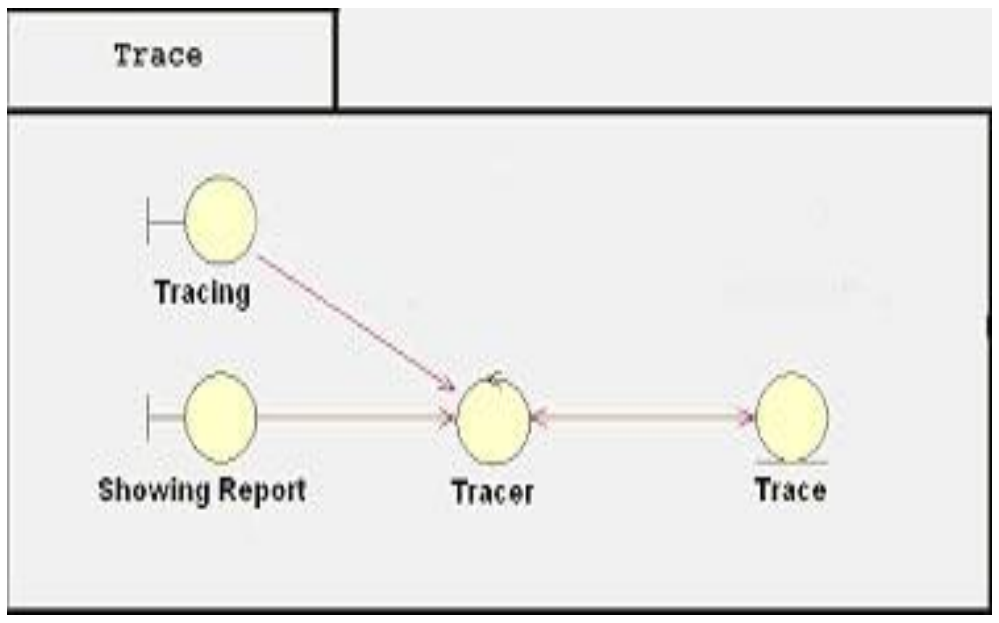

Figure 11: Tracing subsystem.

*2.5 Object Behavior Environment Design*

In the design model, we will adapt and refine the analysis model; the adaptation will be made to the current implementation environment.

*2.5.1 Interaction diagrams*

The system is ready to describe how the different objects are interacting with each other in the Use-cases. We will draw an interaction diagrams (Sequence diagrams) for the main use cases.

The sequence diagram that describes the flow of events for "Open Project" use case is shown in figure 12.

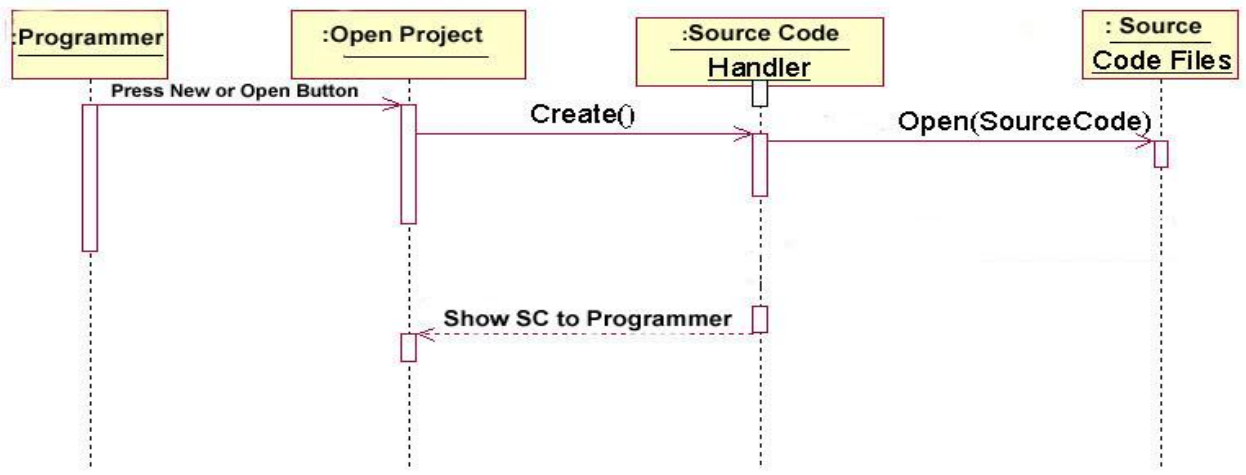

Figure 12: sequence diagram for open project use case.

The sequence diagram that describes the flow of events for "Draw" use case is shown in figure 13.

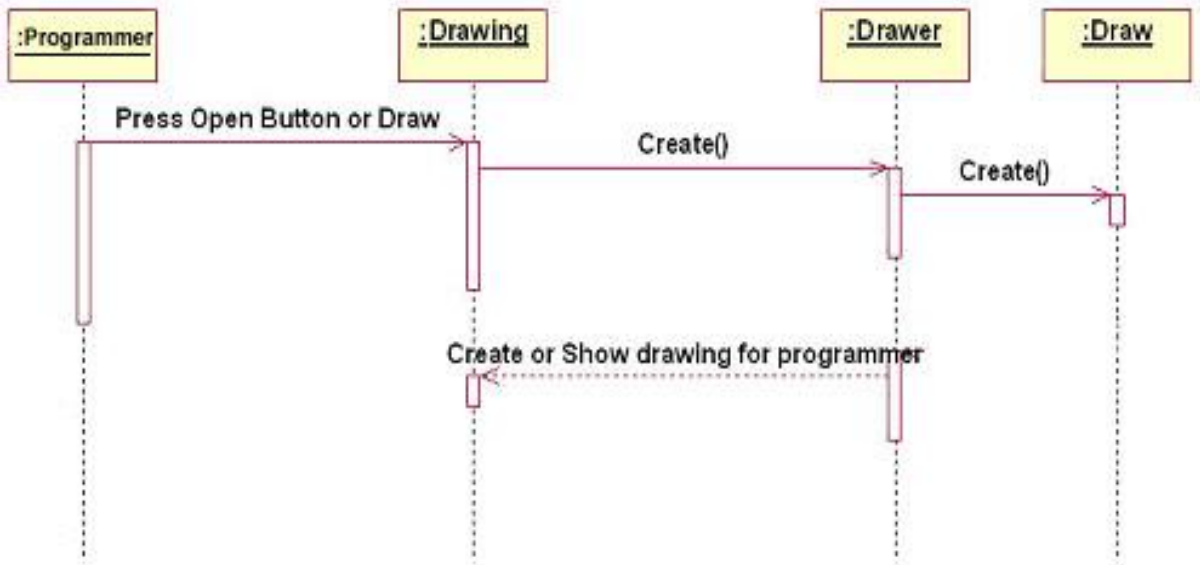

Figure 13: sequence diagram for draw use case.

The sequence diagram that describes the flow of events for "Transformation" use case is shown in the figure 14.

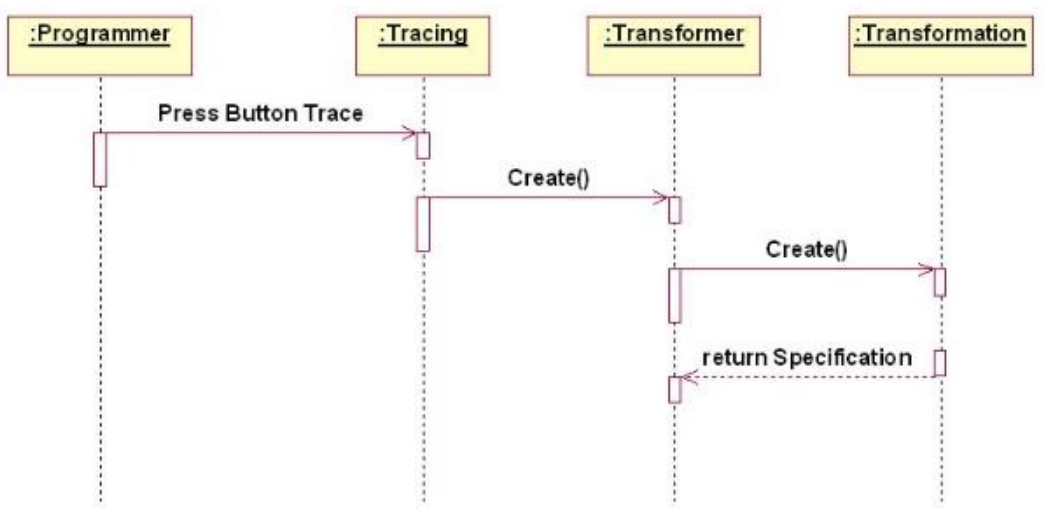

Figure 14: sequence diagram for Transformation use case.

The sequence diagram that describes the flow of events for the whole project is shown in the figure 15.

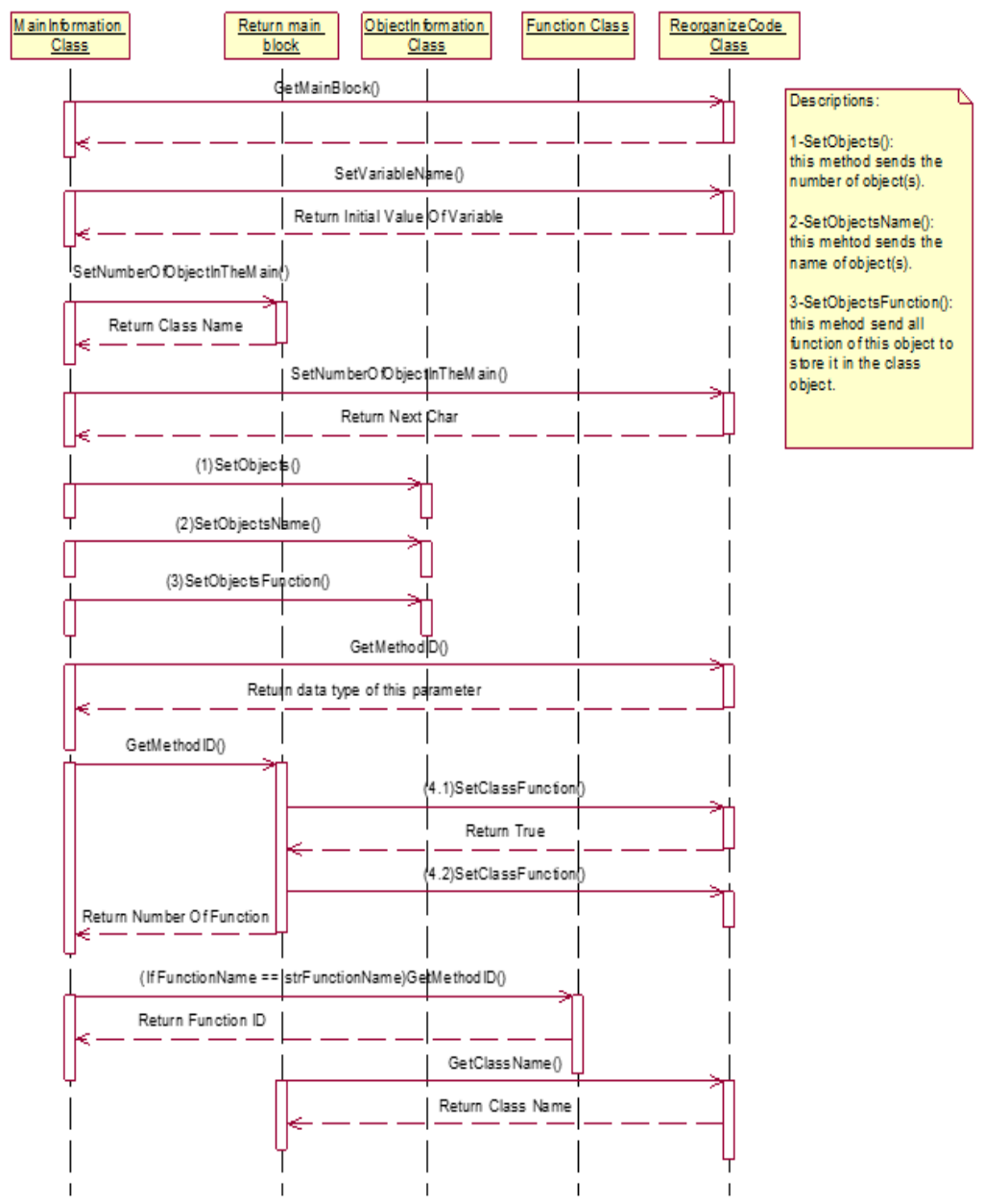

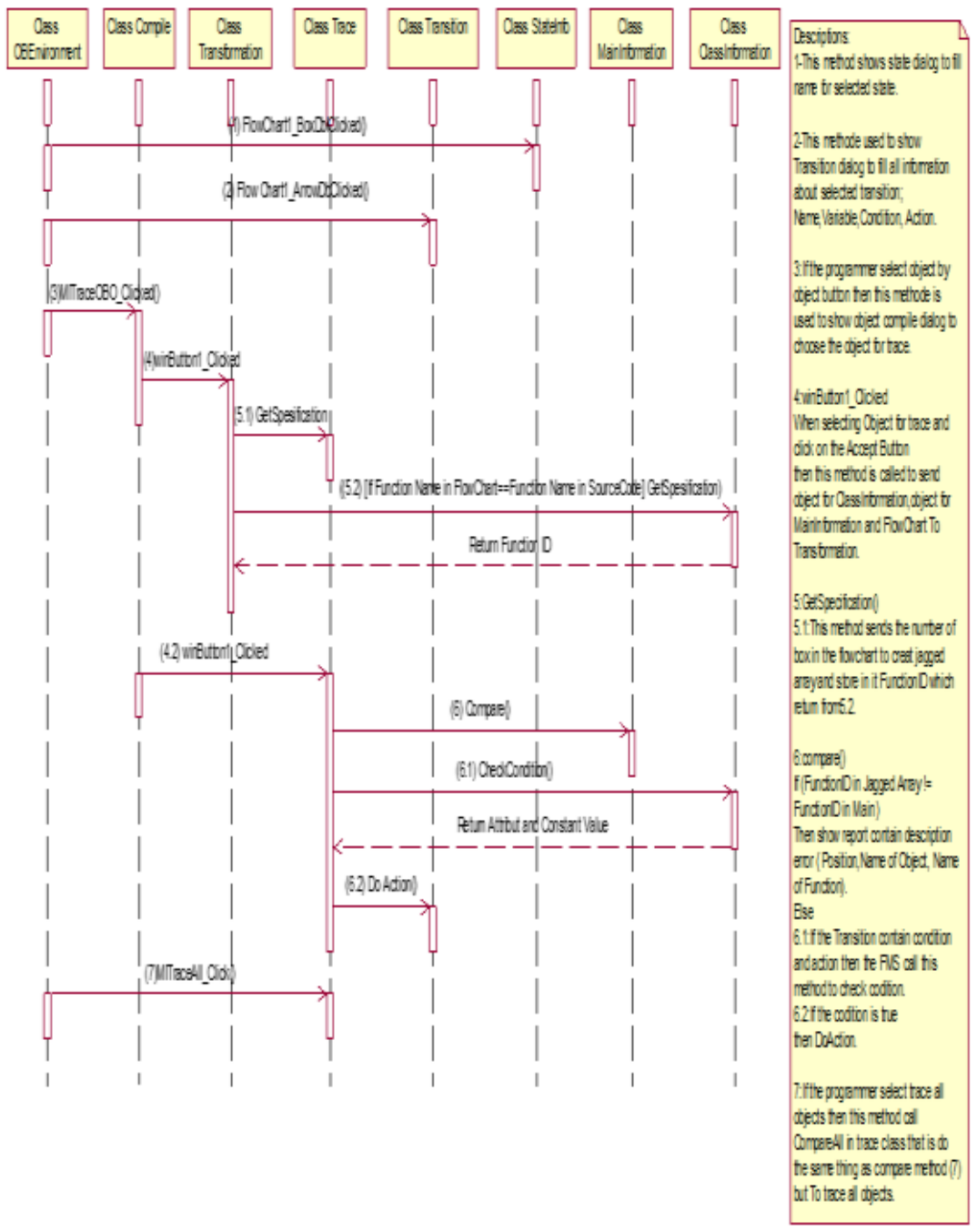

Figure 15: sequence diagram for whole project.

## 2.5.2 *Object Behavior "Activity Diagram"*

The object behavior is described to provide a simplified description to increase the understanding of the block.

- Object behavior for the whole project is shown in figure 16:

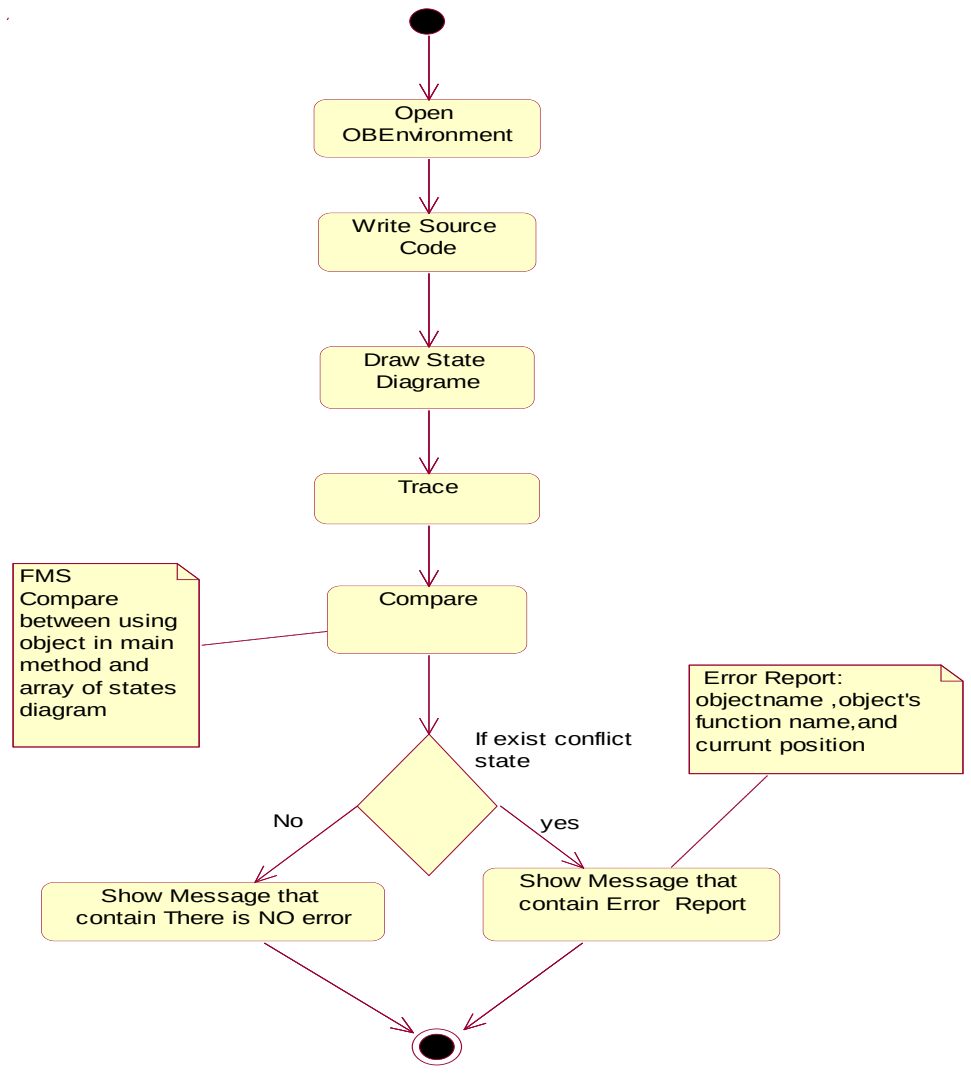

Figure 16: Object behavior for the whole Project.

- Object behaviour for creating or opening project is shown in figure 17:

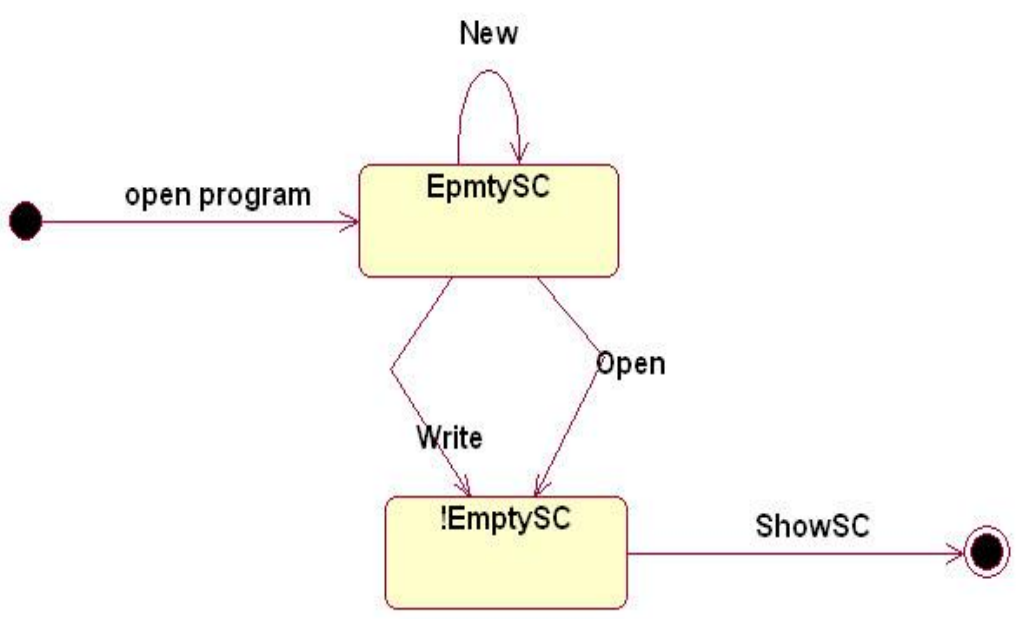

Figure 17: Object behaviour for creating or opening project.

- Object behaviour for Drawing is shown in figure 18:

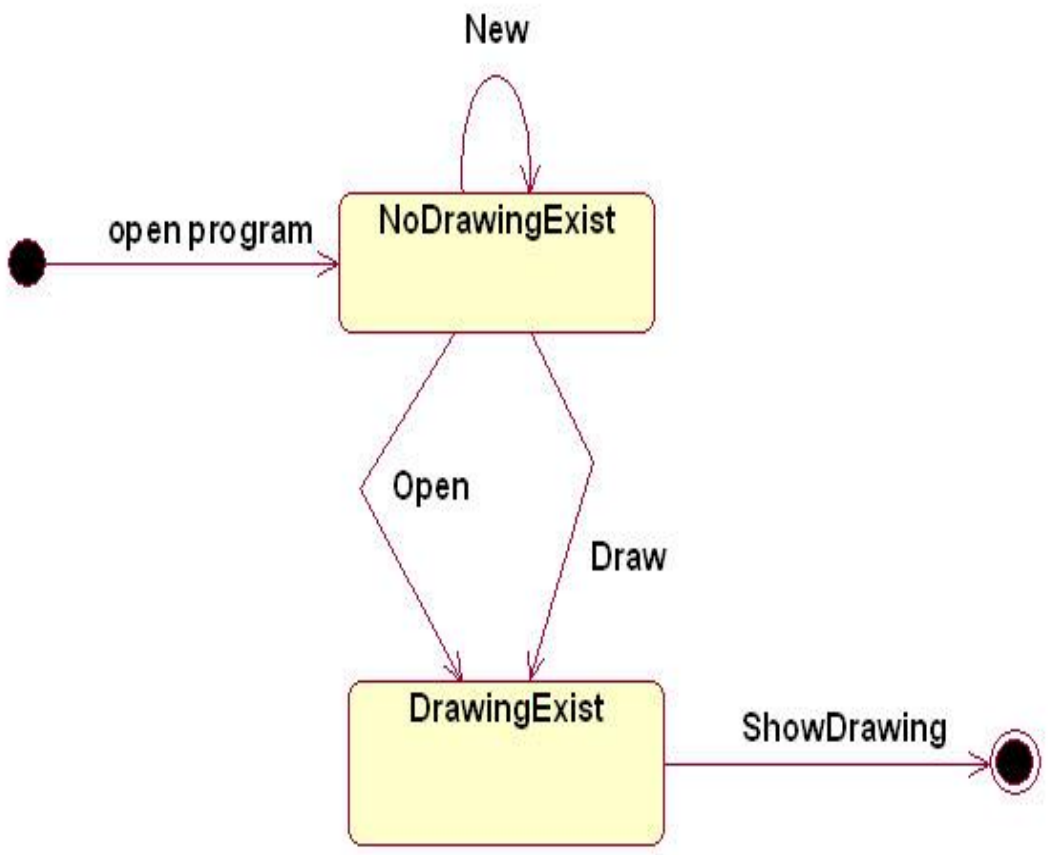

Figure 18: Object behaviour for drawing.

- Object behaviour for Tracing is shown in figure 19:

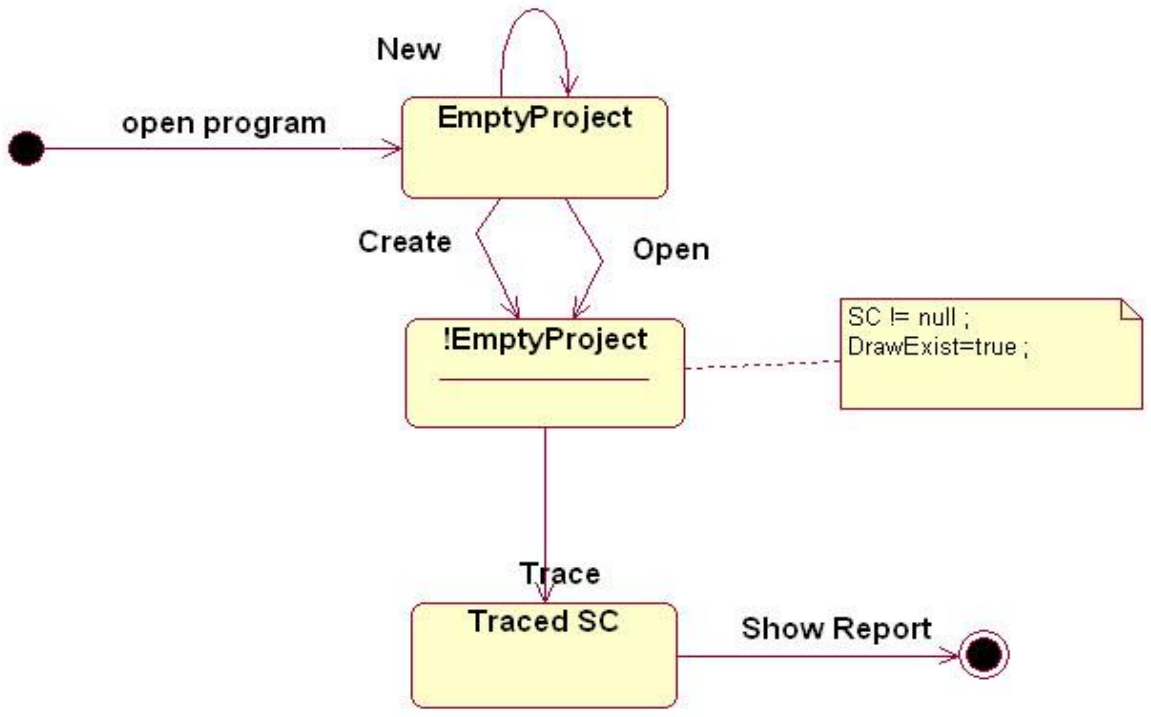

Figure 19: Object behavior for Tracing.

The interfaces objects that are implemented in the OBEnvironment.

### *2.6 Testing Model*

Testing is an activity that aims to investigate the correctness of the system being developed. The testing model includes the following:

• Verification: Checks if the system is built correctly.

• Validation: Checks if the system being built is the right one.

This process is achieved using different levels of testing:

• Unit Testing: Where one unit is tested at a time, in the proposed system the unit is a class.

• Integration Testing: Where we check if the units work together correctly.

• System Testing: Where we check if the whole system performs its intended task as required in the requirements document. [11].

Figure 20 shows the system GUI.

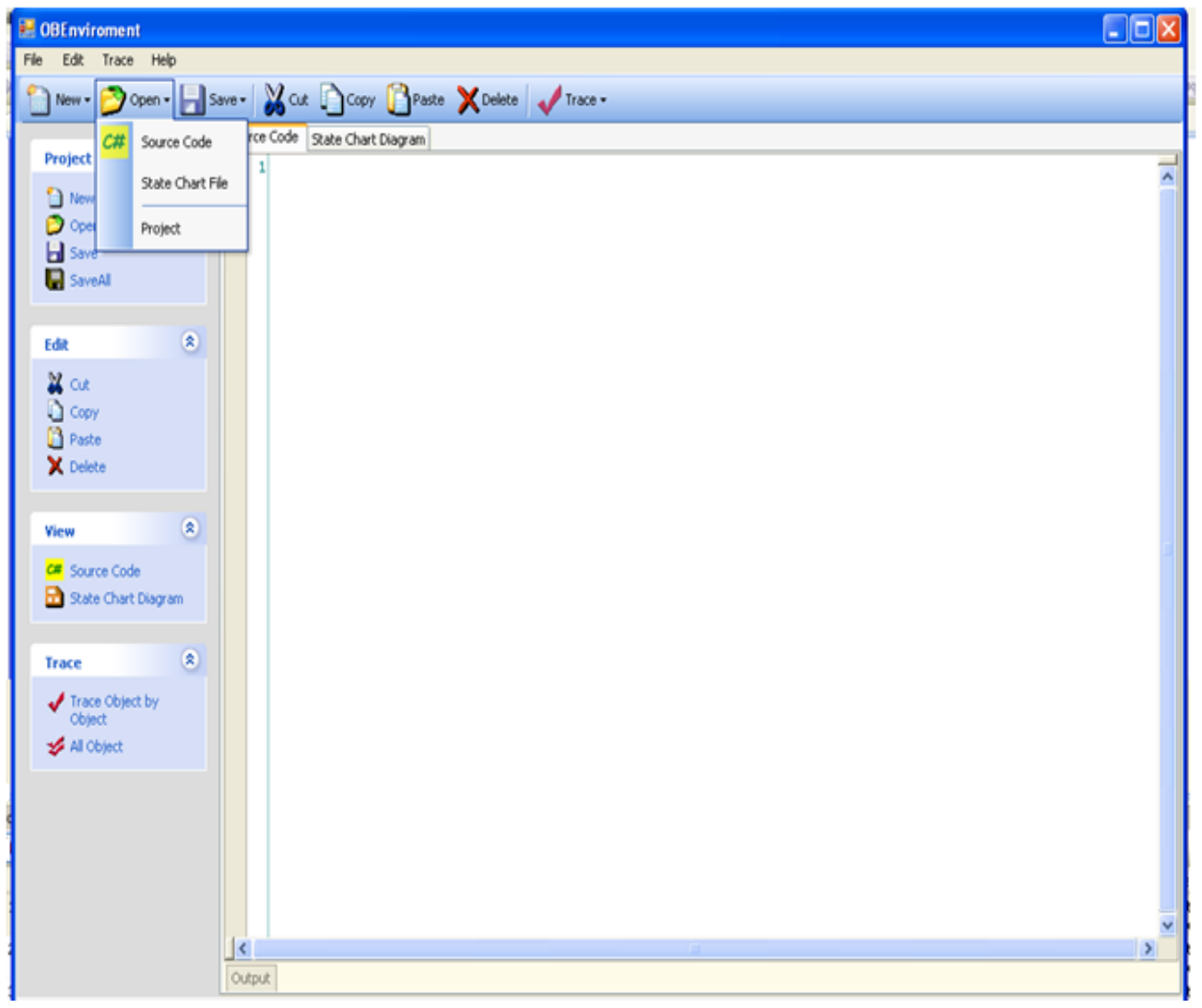

Figure 20: Base Interface.

### *2.6.1 Unit Testing*

Since we are using an Object-Oriented Approach, in other words, Classes then we are concerned with several types of unit testing. Such as:

#### *2.6.1.1 Specification Testing (Black Box Testing)*

The purpose of this type of testing is to test whether the internal building was done successfully. It would be better if all the possible combinations of parameters, variables, and paths in the call are covered, but this is almost always impossible since there are an enormous number of test cases. [11].

#### *2.6.1.2 Structural Testing (White Box Testing)*

The purpose of structural testing is to test the internal structure is correct.

Structural testing is sometimes called program based testing, white box testing or glass box testing; this means that the programmer uses his knowledge of how the unit is implemented when it is tested. [11].

In the proposed system, an example will be taken to demonstrate the available test paths that may be tested in one single class to ensure that all class have been tested individually:

*Class Reorganize_Code: The OBEnvironment reorganize the source code into a standard form for our purposes to help in getting information about the objects. The programmer cannot trace the object behaviour unless the OBEnvironment reorganizes the code, so this is a base class for our system.

To test the OBEnvironment class, Expresso Software (ref) was used to test it step by step. Expresso Software helped to check if the class returns the right value.

This class contains many methods that were tested individually, it arranges the source code (class and the application that uses) and used to get class, main block with it is index, data type of the variables from the main block, to check if the line is a Function or not, and so on.

*Class function & object: These two classes contain many methods that were tested to return important data from the source code. For example, class Function returns the function name, type from the signature; parameters passed to the method. Class object is used to set or get the number of methods used by the object in the main block, to store data of the methods used by the object in the main, return the name, the line of method with the specified ID function. Then those two classes are used to prepare data for tracing process.

*Class MainInformation & ClassInformation: as mentioned before those two classes were used to store and return a specific data for the Trace class. The Transformation and MainInformation class were created to store all variables, all objects name, and all object's functions defined in the Main Block.

*Class Information is used to store all attributes, and all the functions defined in the Class Block, and Get attribute value of existing name, or Return attribute name. This class were tested many times to make sure that it returns the right values.

*Class Transformation: This is one of the most important classes in the proposed system because was used to transform the state diagram into specification. This class tested many different diagrams using jagged array to store object in source code.

*Trace class: This class use the specification to trace the object behaviour, this class is initiated by the programmer which orders the OBEnvironment to trace the Object Behaviour. The OBEnvironment compares the specification with the use of the object in the application, then shows the information about the Objects Behaviour that resulted from the tracing (logical error if existed).

The trace to the Object Behaviour, to check if the condition is true or not, the action of each transition were applied, and compare all object or compare one object that the programmer chose this met.

*Class OBEnvironment: This class contains the interfaces objects that are implemented in the OBEnvironment: Open Project, Drawing, and Trace. Any button in this environment has been tested many times to work in a correct way.

### *2.6.2 Integration Testing*

It decides whether the components of the system are working together properly, this includes testing of blocks, service packages, use cases, subsystems, and the entire system. All the previous classes were tested together several times.

It should be noted that testing is a time-consuming phase of development [11].

### *2.6.3 System Testing*

This process is divided into the following sections:

- Operation Tests.
- Full-Scale Tests.
- Tests of user documentation.

These activities were performed all along with the development of the system [12].

## III. CONCLUSION & FUTURE WORK

### *3.1. Conclusion*

Logical errors are an important topic for research that concerns many software engineers and software companies because of the high maintenance costs, effort and time. A lot of the research papers were written about this topic but there weren't any products produced that deals with the logical errors problem. So, this paper produced an environment that deals with some logical errors that are resulted from the misuse of the objects in the application.

This paper managed to develop an environment that can detect the logical errors that are related to the misuse of the objects in the application this is done by tracing the behaviour of the objects and compare them to some state diagrams of the object class. Here it is assumed that the state chart is the correct object behaviour despite that it might be wrong.

### ***3.2. Future Work***

The proposed system can be improved to produce a new environment that can deal with all kinds of logical errors for OO applications. This would lead to reduce the maintenance cost and increase the software quality.

Create and develop an Object Behaviour Environment that can deal with a dynamic data entry and can take guard conditions from any type and support any object-oriented language.

## ACKNOWLEDGMENT

This research is funded by the Deanship of Research and Graduate Studies in Zarqa University /Jordan.

## APPENDIX

GUI for the proposed system is shown in the following figures.

Create or add project by several methods including:

1) Click the File menu and choose New Project, this allows the programmer to enter a new project in the screen above.

2) Click the File menu and choose Open Project, this allows the programmer to open previously saved project in the screen above.

Create or add source code by several methods including:

1) Click the File menu and choose new source code, this allows the programmer to enter new project in the screen above

2) Click the File menu and choose open project, this allow you to enter new source code in the screen above

3) Click the source code tab appear in the screen above

Create or add state chart diagram by several methods including

1) Click the File menu and choose new statechart diagram this allows the programmer to enter new project in the screen above

2) Click the File menu and choose open project this allows the programmer to enter open previously state statechart diagram in the screen above

You can trace object by object by several methods including

1) Click the list of trace and choose trace object by object as described in the screen below

2) Click the trace object by object button in tool bar appear in the screen below

Here the programmer can choose the object to trace.

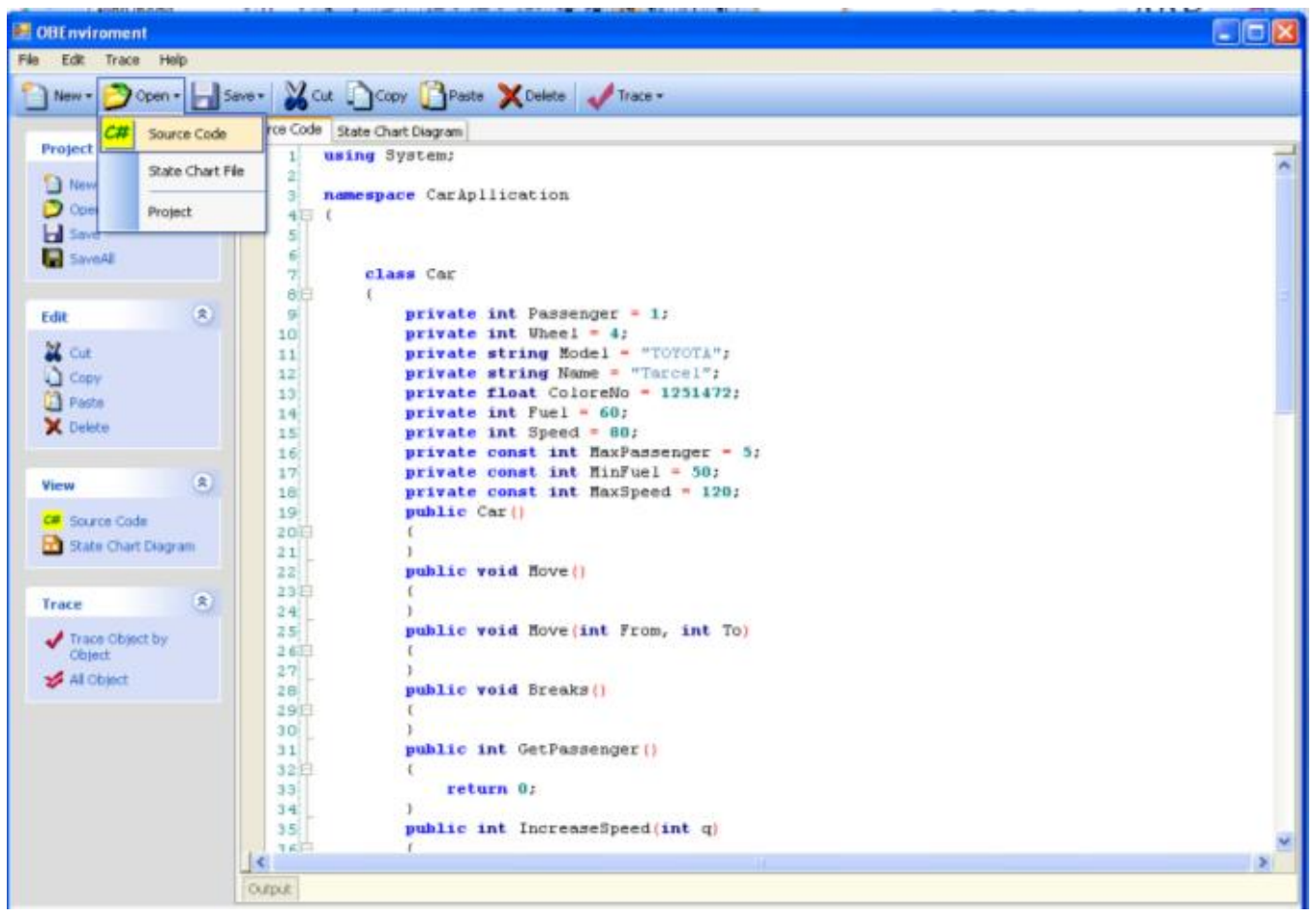

Figure 21: Open source code.

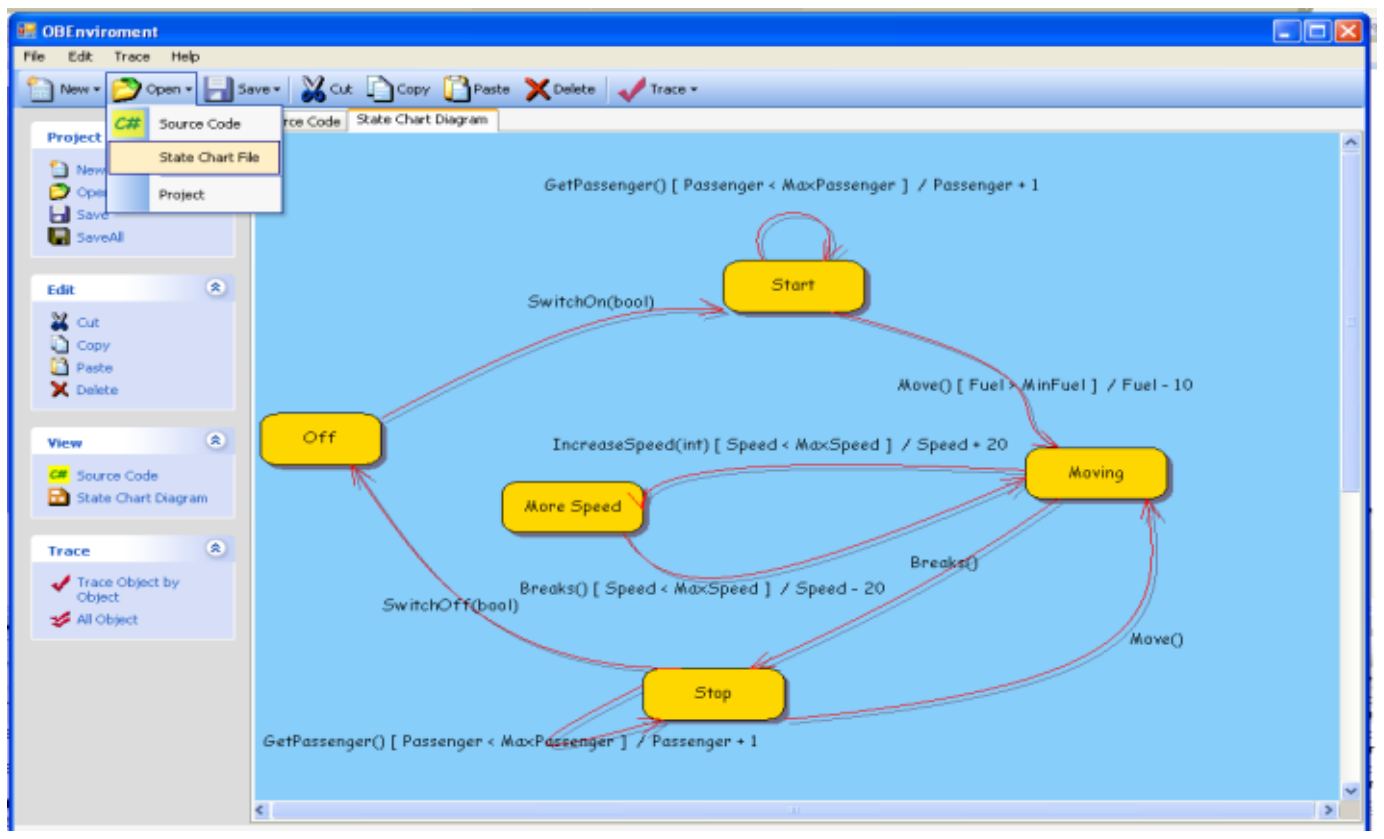

Figure 22: Open state chart diagram.

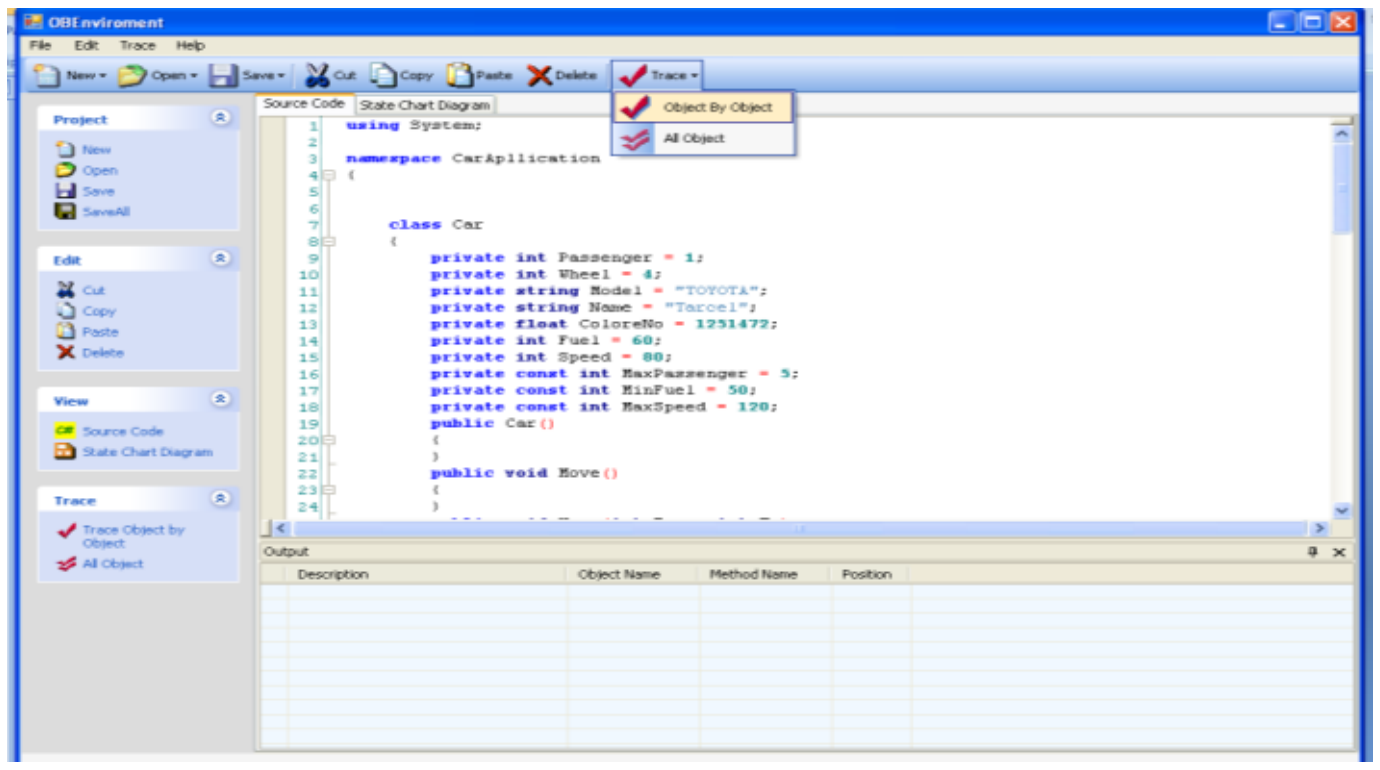

Figure 23: Trace interface object.

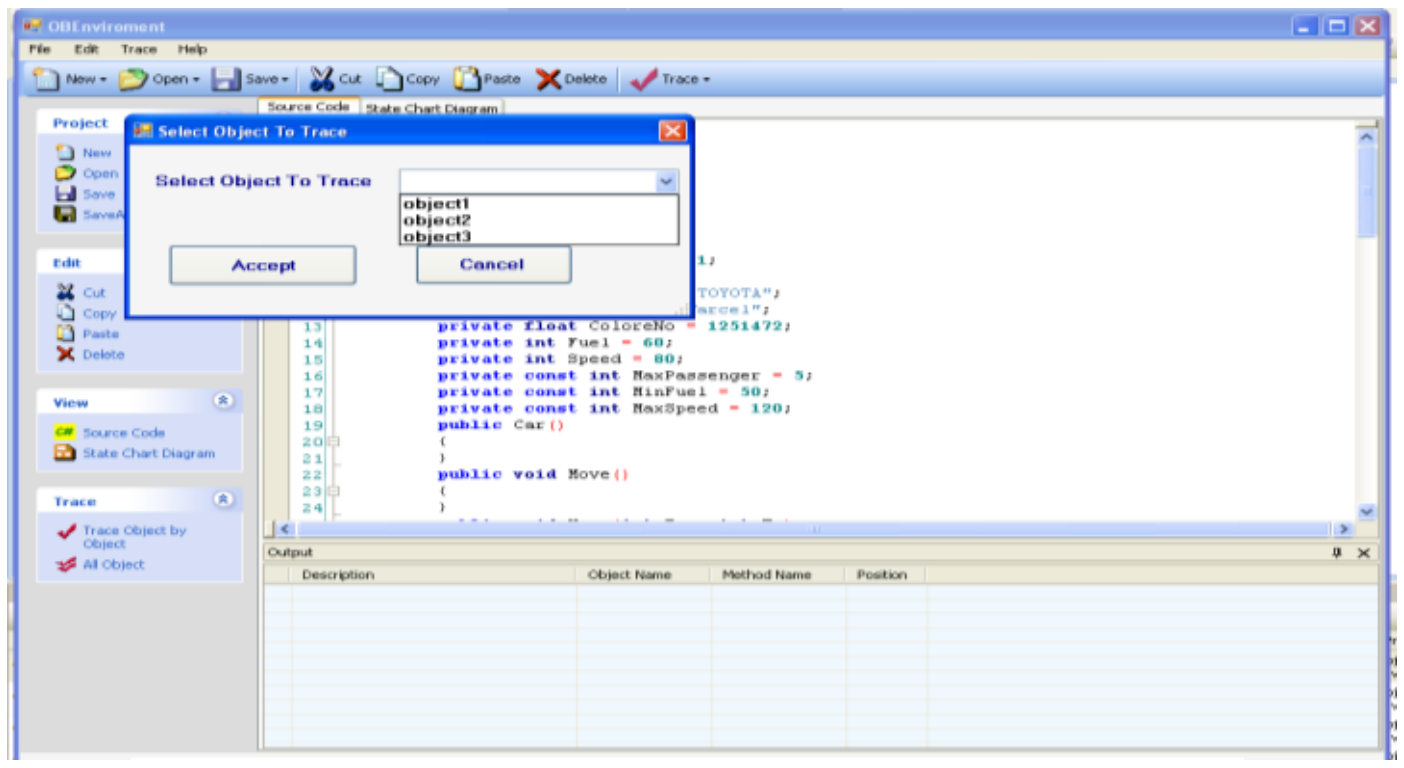

Figure 24: chose object to trace.

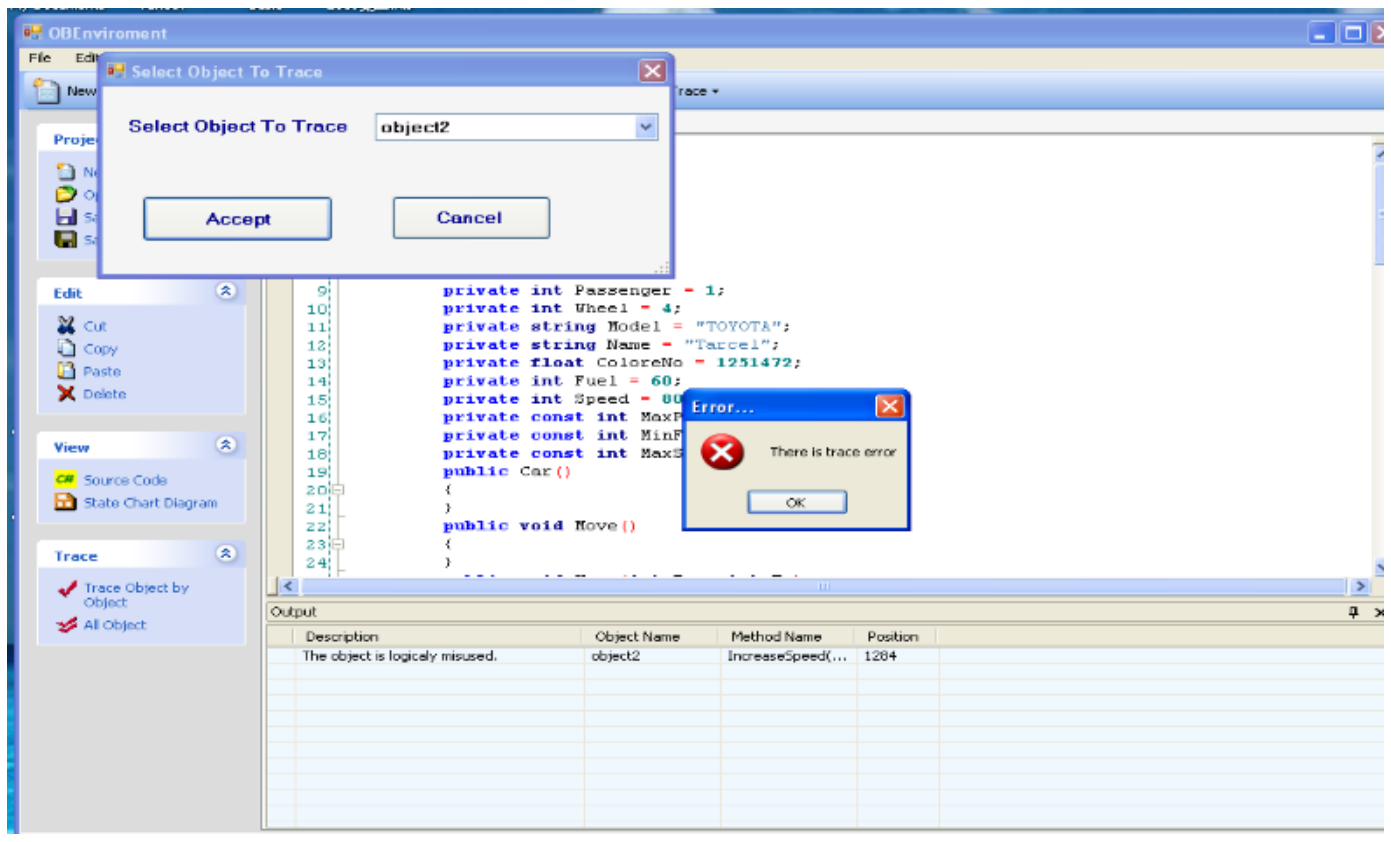

Figure 25: show error message.

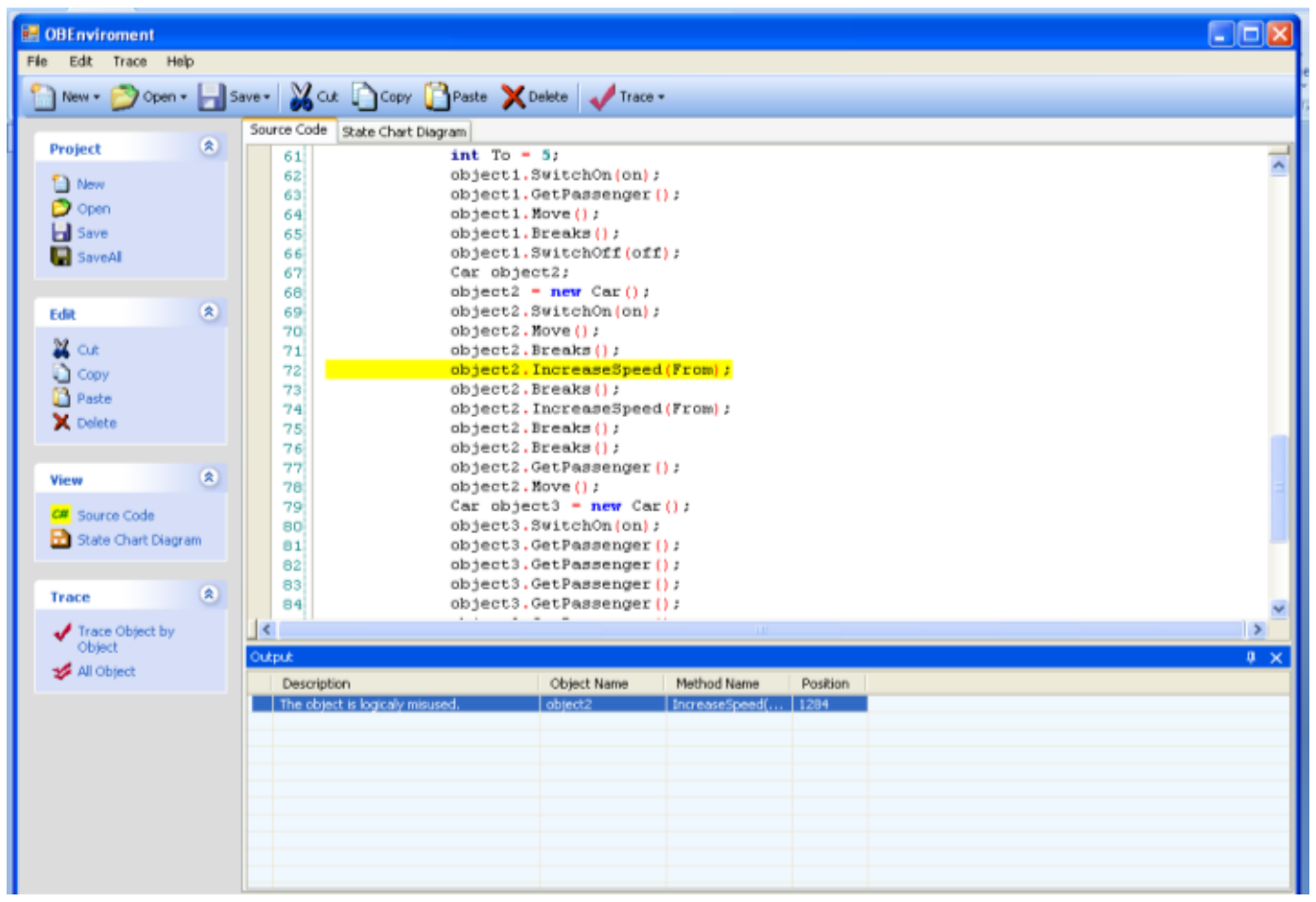

Figure 26: show where error placed.